\documentclass[showpacs,aps,prb,twocolumn,superscriptaddress,10pt]{revtex4-2}
\usepackage{graphicx} % Include figure files
\usepackage{bm}% bold math
\usepackage{color}

\usepackage{amsmath}
\usepackage{amssymb}
\usepackage{enumerate}
\usepackage{xspace}
\usepackage{mathrsfs}
\usepackage{mathptmx}
\usepackage{gensymb} % for degree symbol via \degree
\usepackage[colorlinks=true,linkcolor=blue,citecolor=blue,urlcolor=blue]{hyperref}
\usepackage[utf8]{inputenc}
\usepackage{orcidlink}

\newcommand{\ku}{\ensuremath{k_{\mathrm{u}}}\xspace}
\newcommand{\smsz}{\ensuremath{S_z}}
\newcommand{\smJij}{\ensuremath{J_{ij}}\xspace}

\newcommand{\smB}{\ensuremath{\mathbf{B}}\xspace}
\newcommand{\smBapp}{\ensuremath{\mathbf{B}_{\mathrm{app}}}\xspace}
\newcommand{\sms}{\ensuremath{\mathbf{S}}\xspace}
\newcommand{\bM}{\ensuremath{\mathbf{m}}\xspace}
\newcommand{\bp}{\ensuremath{\mathbf{p}}\xspace}

\newcommand{\bBeff}{\ensuremath{\mathbf{B}_{\mathrm{eff}}}\xspace}
\newcommand{\bsigma}{\ensuremath{\boldsymbol{\sigma}}\xspace}
\newcommand{\thetaSH}{\ensuremath{\theta_\mathrm{SH}}}

\newcommand{\Bstr}{\ensuremath{B_{\mathrm{RT}}^{\mathrm{STT}}}}
\newcommand{\Bstp}{\ensuremath{B_{\mathrm{PT}}^{\mathrm{STT}}}}
\newcommand{\Bsor}{\ensuremath{B_{\mathrm{RT}}^{\mathrm{SOT}}}}

\newcommand{\vampire}{\textsc{vampire} }

\newcommand{\new}[1]{\textcolor{black}{#1}}

\begin{document}

%\title{Spin transfer, spin orbit and spin transport torques in atomistic spin models}
\title{Spin-transfer and spin-orbit torques in the Landau-Lifshitz-Gilbert equation}

\author{Andrea~Meo\orcidlink{0000-0003-2759-4454}}
\affiliation{Department of Physics, University of York, York, YO10 5DD, UK}
\affiliation{Department of Physics, Mahasarakham University, Mahasarakham, 44150, Thailand}
\author{Carenza~E.~Cronshaw\orcidlink{0000-0001-8188-1556}}
\affiliation{Department of Physics, University of York, York, YO10 5DD, UK}
\author{Sarah~Jenkins\orcidlink{0000-0002-6469-9928}}
\affiliation{Department of Physics, University of York, York, YO10 5DD, UK}
\affiliation{TWIST Group, Institut f\"ur Physik, Johannes Gutenberg-Universit\"at, 55128 Mainz}
\author{Amelia~Lees\orcidlink{0000-0001-5576-3978}}
\affiliation{BT Applied Research, Adastral Park, Martlesham, Suffolk, IP5 3RE}
\author{Richard~F.~L.~Evans\orcidlink{0000-0002-2378-8203}}
\affiliation{Department of Physics, University of York, York, YO10 5DD, UK}
\email{richard.evans@york.ac.uk}
\begin{abstract}
Dynamic simulations of spin-transfer and spin-orbit torques are increasingly important for a wide range of spintronic devices including magnetic random access memory, spin-torque nano-oscillators and electrical switching of antiferromagnets. Here we present a computationally efficient method for the implementation of spin-transfer and spin-orbit torques within the Landau-Lifshitz-Gilbert equation used in micromagnetic and atomistic simulations. We consolidate and simplify the varying terminology of different kinds of torques into a physical action and physical origin that clearly shows the common action of spin torques while separating their different physical origins. Our formalism introduces the spin torque as an effective magnetic field, greatly simplifying the numerical implementation and aiding the interpretation of results. The strength of the effective spin torque field unifies the action of the spin torque and subsumes the details of experimental effects such as interface resistance and spin Hall angle into a simple transferable number between numerical simulations. We present a series of numerical tests demonstrating the mechanics of generalised spin torques in a range of spintronic devices. This revised approach to modelling spin-torque effects in numerical simulations enables faster simulations and a more direct way of interpreting the results, and thus it is also suitable to be used in direct comparisons with experimental measurements or in a modelling tool that takes experimental values as input.
\end{abstract}

%\pacs{75.10.Hk,75.20.-g,75.50.Ss,75.60.Jk,75.78.Jp}
\maketitle

\section{Introduction}
Spin-torques describe the action of incoming itinerant electrons on localised magnetic moments in a magnetic material that was first proposed by Slonczewski~\cite{Slonczewski1996} and Berger~\cite{BergerPRB1996}, and are critical to the operation of a wide range of spintronic devices, from spin-transfer torque magnetic random access memory (STT-MRAM)~\cite{Ikeda2010a}, racetrack memory~\cite{Parkin2008}, to spin-orbit torque switching of antiferromagnets~\cite{Wadley2016}. While the fundamental principles of spin-transfer and spin-orbit torques are understood~\cite{Ralph2008}, a wide variety of physical effects dependent on device geometry, interface roughness, temperature, and composition have led to complexity in the interpretation and measurement of switching dynamics. Numerical simulations at micromagnetic~\cite{You2012,Vansteenkiste2014,Abert2018a} and atomistic~\cite{vampire,ChureemartPRB2011,Chureemart2015} length-scales provide further insight compared to purely theoretical calculations~\cite{ZLFPRL2002} allowing for a detailed investigation of nanoscale magnetization dynamics. The standard implementation of spin transfer torques follows the approach of Slonczewski~\cite{Slonczewski1996,Ralph2008}, adding a direct torque into the standard Landau-Lifshitz(LL) or Landau-Lifshitz-Gilbert (LLG) equations, often called the Landau-Lifshitz-Gilbert-Slonczewski (LLGS) equation. While this form directly represents the action of the torque in the equation of motion of the magnetization, it complicates the numerical implementation by requiring an explicitly different equation of motion compared to the standard Landau-Lifshitz-Gilbert equation, or unnecessary computations if the torques are set to zero, slowing down the calculation. The situation is further complicated by the inclusion of adiabatic and non-adiabatic spin torque terms which give non-trivial cross-product expressions when expanding the components in the Gilbert form of the Landau-Lifshitz equation~\cite{Ralph2008}. A more standard approach when considering magnetic interactions in micromagnetic or atomistic simulations is to write an effective magnetic vector field that acts on the magnetization. This avoids the need to add additional torques for each and every magnetic interaction, simplifying the numerical implementation and improving computational performance. The general structure of spin-torques means it is not straightforward to implement as an effective magnetic field, and adding the torque terms directly \cite{TorrejonPRB2015,MeoPRB2021,GarciaSanchez2020} leads to a cross-pollution of the adiabatic and non-adiabatic terms, making it difficult to isolate the role of the different spin-torque component. Here we present a formulation of spin-torques as effective magnetic fields, greatly simplifying the numerical implementation and providing an intuitive understanding of the action of different spin-torques. 

\section{Theory}
The literature describing the general action of spin torques uses a wide range of terminology describing a combination of physical effects, actions of the torques or named after the inventor. The most commonly named torques are spin transfer torque (STT) and spin orbit torque (SOT), alluding to their physical origins\cite{Brataas2017}. Depending on what is expected to be the dominant effect, STT is also described as damping-like (DL) torque, while (SOT) is often described as a field-like (FL) torque, leading to magnetization precession around the itinerant spin polarization. In reality both spin-transfer and spin-orbit torques can lead to both effects, with the relative importance dependent on the particular geometry and material properties of the device. Furthermore the term field-like is ambiguous in that magnetic fields cause both damping and precessional effects. In Tab. 1 we propose a rationalisation of terminology to make the physical origin \new{(from spin-transfer effects or spin-orbit effects)} and action \new{(pure precession and pure relaxation)} of the different torques clear. This taxonomy explicitly includes the physical origin of the torque (from spin transfer and spin-orbit origins) as well as the action of the torque on the magnetization, causing it to either precess around the spin polarization direction $\bp$ or relax towards it. \new{Compared to existing descriptions in the literature this is more straightforward and using consistent language so that the physical action of the torque is explicitly stated. A similar approach can be used when describing torques arising from spin pumping~\cite{Brataas2017} and in more advanced models of spin transport that explicitly calculate the spin-accumulation and spin-current.}

\begin{table}[!tb]
\centering % used for centering table
\begin{ruledtabular}
\begin{tabular}{l l}
New terminology & Old terminology \\
\hline % inserts single horizontal line
spin transfer relaxation torque & spin transfer torque\\
(STRT)                          & damping-like torque\\
spin orbit relaxation torque    & adiabatic spin torque\\
(SORT)                          & Slonczewski torque\\
\hline % inserts single horizontal line
spin transfer precession torque & spin-orbit torque\\
(STPT)                          & field-like torque\\
spin orbit precession torque    & non-adiabatic spin torque\\
(SOPT)                          & 
\end{tabular}
\end{ruledtabular}
\caption{Taxonomy of spin torques. To simplify the nomenclature we define torques in terms of their action on the magnetization (relaxation or precession) and their physical origin (spin-transfer and spin-orbit). The synonymous terms are listed on the right hand side.}
\label{tab:params}
\end{table}

Having defined the description of the torques, numerically we only need to consider two effects. The first is a torque that a causes a purely precessional motion of the magnetization, and the second is a torque that causes a pure relaxation of the magnetization. Conventionally when considering spin transfer torques, one often writes the form of the Landau-Lifshitz equation augmented by the Slonczewski spin transfer torque \cite{Slonczewski1996}:
\begin{equation}
    \frac{\partial \bM}{\partial{t}} = -\gamma_e \bM \times \smB + \alpha_G \bM \times\frac{\partial \bM}{\partial t} -\gamma_e H_s \bM \times (\bM \times \bp)
\label{eq:LLS}
\end{equation}
where $\bp$ is the spin polarization and $H_s$ represents the strength of the field associated to STT. \new{Eq.~\ref{eq:LLS} is an implicit equation, where the term $\partial \bM / \partial{t}$ appears on both sides of the equation. This equation may be expressed in explicit form after some manipulation~\cite{Melcher2013} which introduces terms with both adiabatic $\bM \times (\bM \times \bp)$ and non-adiabatic $\bM \times \bp$ symmetry but with prefactors that depend on the Gilbert damping $\alpha_{\mathrm{G}}$ ~\cite{Brataas2017}. The introduction of terms of different symmetry and implicit dependence on the Gilbert damping is a non-obvious consequence of moving from the implicit Landau-Lifshitz-Gilbert-Sloczewski equation to explicit form and often not discussed when performing numerical simulations~\cite{BeikMohammadi2021,Taniguchi2018,Pathak2020,Zhuo2022}, despite numerical packages solving the explicit form~\cite{Vansteenkiste2014,You2012}. This complexity contributes to the confusion in the literature and also difficulty when interpreting experimental measurements between materials with different Gilbert damping constants. }

\section{Methodology}
To simplify the understanding of the effects of relaxational and precessional torques and their computational implementation, we apply them as conventional effective magnetic fields within the usual Landau-Lifshitz-Gilbert (LLG) equation. 

\subsection{Derivation of spin-transfer-torque fields}
In the following we derive the expression for the LLG equation in the presence of STT fields in its solved form, i.e. where time derivatives of $\bM$ appears on one side of the equation only. Let us consider the LLS equation:
\begin{equation}
    \label{eq:LLG_STT_Slonc}
    \frac{\partial \bM}{\partial t} = -\gamma_e \bM \times \smB + \alpha_G \bM \times\frac{\partial \bM}{\partial t} -\gamma_e H_s \bM \times (\bM \times \bp) \, .
\end{equation}
To transform the equation into its explicit formulation we can expand the term $\alpha_G \bM \times\frac{\partial \bM}{\partial t}$ on the right hand side. Doing so we obtain the following expression:
\begin{align}
    \label{eq:LLG_STT_step1}
    \alpha_G \bM \times \frac{\partial \bM}{\partial t} & = \\ \nonumber
    - \gamma_e \alpha_G \bM \times \left( \bM \times \smB\right) 
    & + \alpha_G^2 \bM \times \left( \bM \times\frac{\partial \bM}{\partial t} \right) \\ \nonumber
    - \gamma_e \alpha_G H_s \bM & \times \left[ \bM \times \left( \bM \times \bp \right) \right]
\end{align}
which includes the quadruple cross product $\bM \times \left[ \bM \times \left( \bM \times \bp \right) \right]$.
By taking $\bM \times \bp$ as a vector and not solving for it, the quadruple cross product becomes a triple cross product, which can be rewritten exploiting the triple product expansion rule:
\begin{equation}
    \bM \times \left[ \bM \times \left( \bM \times \bp \right) \right] = 
    \bM \left[ \bM \cdot \left( \bM \times \bp \right) \right] - 
    \left( \bM \times \bp \right) \left( \bM \cdot \bM \right) \,.
\end{equation}
The expression can be further simplified by exploiting that $\bM \cdot \left( \bM \times \bp \right) = 0$ and $\bM \cdot \bM = 1$. We obtain the following expression for the quadruple cross product:
\begin{equation}
    \label{eq:quadruple_cross_product}
    \bM \times \left[ \bM \times \left( \bM \times \bp \right) \right] = - \bM \times \bp \, .
\end{equation}
Similarly, the triple cross product $\bM \times \left( \bM \times\frac{\partial \bM}{\partial t} \right)$ can be rearranged, by exploiting that $\bM$ and ${\partial \bM}/{\partial t}$ are orthogonal, into:
\begin{equation}
    \label{eq:triple_cross_product_dm}
    \bM \times \left( \bM \times\frac{\partial \bM}{\partial t} \right) = 
    \bM \left( \bM \cdot \frac{\partial \bM}{\partial t} \right) - \frac{\partial \bM}{\partial t} \left( \bM \cdot \bM \right)
    = - \frac{\partial \bM}{\partial t} 
\end{equation}
By plugging eqns.~\ref{eq:quadruple_cross_product} and ~\ref{eq:triple_cross_product_dm} into eqn.~\ref{eq:LLG_STT_step1}, the expression for $\alpha_G \bM \times\frac{\partial \bM}{\partial t}$ reads:
\begin{align}
    \label{eq:LLG_STT_step2}
    \alpha_G \bM \times \frac{\partial \bM}{\partial t}  =
    - \gamma_e \alpha_G \bM \times \left( \bM \times \smB \right) \\ \nonumber
    - \alpha_G^2 \frac{\partial \bM}{\partial t} 
    + \gamma_e \alpha_G H_s \left( \bM \times \bp \right) \, .
\end{align} 
If we substitute the result just found for $\alpha_G \bM \times \frac{\partial \bM}{\partial t} $ into the original LLS equation, we obtain:
\begin{align}
    \label{eq:LLG_STT_step3}
    \frac{\partial \bM}{\partial t} = 
    - \gamma_e \left( \bM \times \smB \right) \\ \nonumber
    - \gamma_e \alpha_G \bM \times \left( \bM \times \smB \right)  
    - \alpha_G^2 \frac{\partial \bM}{\partial t} \\ \nonumber
    + \gamma_e \alpha_G H_s \left( \bM \times \bp \right)
    - \gamma_e H_s \bM \times (\bM \times \bp) \, .
\end{align}
Finally, by collecting terms in ${\partial \bM}/{\partial t}$ on the left hand side we obtain the solved form of the LLS equation, which is more suitable for efficient computation:
\begin{align}
    \label{eq:LLG_STT_step4}
    \left(1+\alpha_G^2\right)\frac{\partial \bM}{\partial t} = \\ \nonumber
    - \gamma_e \left( \bM \times \smB \right) 
    - \gamma_e \alpha_G \bM \times \left( \bM \times \smB \right)  \\ \nonumber
    + \gamma_e \alpha_G H_s \left( \bM \times \bp \right)
    - \gamma_e H_s \bM \times \left(\bM \times \bp\right) \, .
\end{align}
The terms in $\bM \times \smB$ and $\bM \times \bp$ have similar forms, thus we can define an effective field $\bBeff = \smB + H_s \left(\bM \times \bp\right)$ such that 
\begin{align}
    \label{eq:Beff}
    \bM \times \bBeff &= \bM \times \smB + H_s \bM \times \left(\bM \times \bp\right) \nonumber \\
    \bM \times \left( \bM \times \bBeff \right) &=  \bM \times \left( \bM \times \smB \right) - H_s \left( \bM \times \bp \right) \, .
\end{align}
We note that both components of the term describing the relaxational motion appear with an $\alpha_G$ factor in eqn.~\ref{eq:LLG_STT_step4}, consistently with the LLG formalism.
Thus, if we now substitute eqns.~\ref{eq:Beff} into eqn.~\ref{eq:LLG_STT_step4} rewriting the expression in terms of \bBeff and dividing by $\left(1+\alpha_G^2\right)$, we obtain an expression analogous to the standard explicit LLG equation in terms of conventional fields:
\begin{equation}
    \label{eq:LLG_STT_solved}
    \frac{\partial \bM}{\partial t} = 
    - \frac{\gamma_e}{\left(1+\alpha_G^2\right)} \left( \bM \times \bBeff \right) 
    - \frac{\gamma_e \alpha_G }{\left(1+\alpha_G^2\right)} \left[ \bM \times \left( \bM \times \bBeff \right) \right] \, .
\end{equation}
An analogous derivation can be obtained for the general SOT case by simply replacing the third term on the RHS of eqn.~\ref{eq:LLG_STT_Slonc} with an expression for SOT fields.

\subsection{Spin-transfer torque fields}
Specifying our approach to the application of spin-transfer torque (STT), we can describe the magnetisation dynamics under the effect of STT by adding the following field to the standard LLG equation: 
\begin{equation}
    \label{eq:STT}
    \smB_{\mathrm{STT}} = B_{\mathrm{PT}}^{\mathrm{STT}} \left(\bp - \alpha_G \bM \times \bp\right) + B_{\mathrm{RT}}^{\mathrm{STT}} \left(\bM \times \bp + \alpha_G \bp \right)
\end{equation}
which comprises both the relaxational and precessional components. \new{Here we introduce adiabatic (relaxational) and non-adiabatic (precessional) terms as fuly independent terms rather than coupled terms of the same symmetry that naturally arise from Eq.~\ref{eq:LLS}. This explicitly expresses spin torques without any dependence on the Gilbert damping, which would naturally arise from the explicit form of Eq.~\ref{eq:LLS} making the action of the pure torques clearer. It is possible to return to the LLGS form in Eq.~\ref{eq:LLS} by setting $B_{\mathrm{PT}}^{\mathrm{STT}} = -\alpha_{\mathrm{G}}B_{\mathrm{RT}}^{\mathrm{STT}}$. In this case the action of the STT is always mixed (containing precessional and relaxational components) and dependent on the magnitude of the Gilbert damping, thus complicating the interpretation of the effect of the spin torque in simulations and experiments.}

The strength of the STT terms $B_{\mathrm{RT,PT}}^{\mathrm{STT}}$ depends on the injected areal current density $j_e$ (A/m$^2$) and at the micromagnetic level can be expressed as \cite{Slonczewski1996,Slonczewski2005,Bertotti2005,Serpico2006}:
\begin{align}
    \label{eq:STT_strengthsRT}
    B_{\mathrm{RT}}^{\mathrm{STT}} &= \frac{\hbar \eta j_e}{2e(1+\lambda \bM \cdot \bp) M_s d} \\ 
    \label{eq:STT_strengthsPT}
    B_{\mathrm{PT}}^{\mathrm{STT}} &= \beta_{\mathrm{STT}} \frac{\hbar \eta j_e}{2e(1+\lambda \bM \cdot \bp) M_s d} = \beta_{\mathrm{STT}} B_{\mathrm{RT}} \, ,
\end{align}
where $\eta$ is the spin polarisation, $\lambda$ is the
spin torque asymmetry often taken as $\lambda=\eta^2$, $M_\mathrm{S}$ is the saturation magnetisation and $d$ the thickness of the free layer (FL). The degree of non-adiabaticity of the system \cite{Zhang2004b,Abert2018a} is defined by the $\beta_{\mathrm{STT}}$ is a factor that determines the strength of the precession torque field $B_{\mathrm{PT}}$ in relation to the relaxation torque field $B_{\mathrm{RT}}$. It has been found that in spin vales, a trilayer structure where two ferromagnets are separated by a non-magnetic metal, $B_{\mathrm{PT}}$ is negligible. On the other hand, in magnetic tunnel junctions (MTJs), where the non-magnetic metal of the pin valve is replaced by an insulator, $B_{\mathrm{PT}}$ can be as large as $B_{\mathrm{RT}}$ or even stronger. Typical values of spin transfer torques fields are 0.0001-0.2 Tesla for typical current densities and device geometries. The reformulation of spin-torques as effective magnetic fields now allows a straightforward comparison with other magnetic fields such as coercivity, providing an intuitive understanding of their relative strength and importance.

\subsection{Spin-transfer torque in atomistic models}
In atomistic models compared to micromagnetic calculations the situation becomes slightly more complex due to the discrete nature of atomistic spins, the explicit localisation of spintronic effects on the atomic scale at interfaces~\cite{MeoPRB2021} and high temperatures where the saturation magnetisation $M_\mathrm{S}$ tends to zero, causing a divergence in the effective field. Here the saturation magnetization and thickness are poorly defined and so we can reformulate the effective spin transfer torque field at the atomistic level. In the micromagnetic limit the effective volume of action $V$ is assumed to be defined by the film thickness $d$, typically 1-2 nanometres thick and over a device area $A$. Considering the micromagnetic case in Eq.~\ref{eq:STT_strengthsRT} we have a total moment $M = M_\mathrm{S}V$, film thickness $d$ and current density $j_e$ through an interfacial area $A$ and acting over a volume $V$. We can reformulate Eq.~\ref{eq:STT_strengthsRT} equivalently as
\begin{equation}
    B_{\mathrm{RT}}^{\mathrm{STT}} = \frac{\hbar \eta j_e}{2e(1+\lambda \bM \cdot \bp) M_s d} = \frac{\hbar \eta j_e A}{2e(1+\lambda \bM \cdot \bp) M_s V}
\end{equation}
where the volume of action $V = Ad$. Moving to an atomistic description we now have 
\begin{equation}
    B_{\mathrm{RT}}^{\mathrm{STT}} = \frac{\hbar \eta j_e A}{2e(1+\lambda \bM \cdot \bp) N\mu_{\mathrm{S}}}
\end{equation}
where N = $V/V_{\mathrm{at}}$ is the number of atoms,  
$\mu_\mathrm{S}$ is the local atomic moment with atomic magnetic volume $V_{\mathrm{at}}$ defined within its unit cell as
\begin{equation}
    V_{\mathrm{at}} = \frac{a^3}{N_{\mathrm{uc}}}
\end{equation}
where $a$ is the unit cell size and $N_{\mathrm{uc}}$ is the number of magnetic atoms per unit cell, for example $N_{\mathrm{uc}}=1$ for simple cubic, $N_{\mathrm{uc}}=2$ for body-centred cubic and $N_{\mathrm{uc}}=4$ for face-centred cubic crystals. The effective volume $V$ relevant for spin-torque effects is important as it determines the strength of the spin-torque field. The localisation of spin-torque effects is a complex topic and strongly material dependent~\cite{PhysRevB.66.014407,GalantePRB2019}, and correctly parameterising the atomistic description of spin-transfer torque requires an assumption of how far incident spin currents penetrate into a ferromagnet. The two simplest approximations here that the currents penetrate the entire thickness of the ferromagnetic layer (assuming a thin film of a few nanometres), or that the spin current is absorbed in an interfacial monolayer. In the latter case we can compute the effective volume straightforwardly considering a single atom per unit cell at the interface, so that $A = a^2$ and $N = 1$, such that 
\begin{equation}
    B_{\mathrm{RT}}^{\mathrm{STT}} = \frac{\hbar \eta j_e a^2}{2e(1+\lambda \bM \cdot \bp) \mu_{\mathrm{S}}}
\end{equation}
with the equivalent expression for the precessional torque magnitude. This expression avoids the divergence in the micromagnetic form as the magnetisation tends to zero near the Curie temperature and is suitable for atomistic simulations. Finally when considering temperature effects from the atomistic approach the spin-transfer torque field acting on all spins is temperature independent in this formalism and depends only on the effective angular momentum transfer from the incoming spin current.

\subsection{Spin-orbit torque fields}
Spin-orbit torques (SOT) can be described by using an analogous formalism to spin-transfer torque, where \bp is replaced by the spin polarisation unit vector \bsigma. \bsigma represents the direction of the polarisation of the spin current induced by the flow of electrons in a non magnet and it is perpendicular to the electron flow. In our case we discard the component of \bsigma that is normal to the interface between the non-magnet and the magnet, as it would yield negligible spin accumulation.

%\cmr{magnetic forces on the electrons, principle of reciprocity and JS.s represntation.}\comAn{what I refer to is the fact that there is spin accumulation also along the other in-plane direction orthogonal to the current. However, since there is no interface there, the contribution is small.}

The expression for the field for SOT is:
\begin{equation}
    \label{eq:SOT}
    \smB_{\mathrm{SOT}} = B_{\mathrm{PT}}^{\mathrm{SOT}} \left(\bsigma - \alpha_G \bM \times \bsigma\right) + B_{\mathrm{RT}}^{\mathrm{SOT}} \left(\bM \times \bsigma + \alpha_G \bsigma \right) \, .
\end{equation}
The strength of the SOT fields depends on the mechanism involved, such as spin Hall effect (SHE), Rashba effect or inverse spin galvanic effect (iSGE) \cite{Manchon2019b}. For simplicity here we consider SHE as the dominant mechanism yielding spin-orbit torque, but the subtle differences in these different mechanisms are independent of the formalism and can be simply represented by different combinations of relaxation and precession torques. At the micromagnetic level the spin-orbit torque field strengths $B_{\mathrm{RT}}$ and $B_{\mathrm{PT}}$ are given by \cite{Manchon2019b}:
\begin{align}
    \label{eq:SOT_strengths}
    B_{\mathrm{RT}}^{\mathrm{SOT}} &= \frac{\hbar j_e \thetaSH}{2e M_s d} \\
    B_{\mathrm{PT}}^{\mathrm{SOT}} &= \beta_{\mathrm{SOT}} \frac{\hbar j_e \thetaSH}{2e M_s d} = \beta_{\mathrm{SOT}} B_{\mathrm{RT}} \, ,
\end{align}
where $j_e$ is the injected current density, \thetaSH is the spin Hall angle and it gives the conversion efficiency of electrical current into spin current, $d$ is the ferromagnet thickness and $M_s$ is the magnetisation of the ferromagnet. Here $\beta_{\mathrm{SOT}}$ is an empirical scaling factor that relates the strength of the precessional term with the relaxation term. It is in general assumed to be less than one, but in particular systems it could be larger. The equivalent atomistic expressions for a purely interfacial spin-orbit torque acting on a single monolayer of atoms are
\begin{align}
    \label{eq:SOT_strengths}
    B_{\mathrm{RT}}^{\mathrm{SOT}} &= \frac{\hbar j_e \thetaSH a^2}{2e \mu_{\mathrm{S}}} \\
    B_{\mathrm{PT}}^{\mathrm{SOT}} &= \beta_{\mathrm{SOT}} \frac{\hbar j_e \thetaSH a^2}{2e \mu_{\mathrm{S}}} = \beta_{\mathrm{SOT}} B_{\mathrm{RT}} \, .
\end{align}

\section{Results}
Having defined our formalism and numerical implementation we present a series of sample calculations showing the intrinsic effects of relaxational and precessional spin torques as well as well as some topical examples of application of spin torques.
\subsection{Intrinsic dynamics of spin torques}

\begin{figure}
    \centering
    \includegraphics[width=\columnwidth]{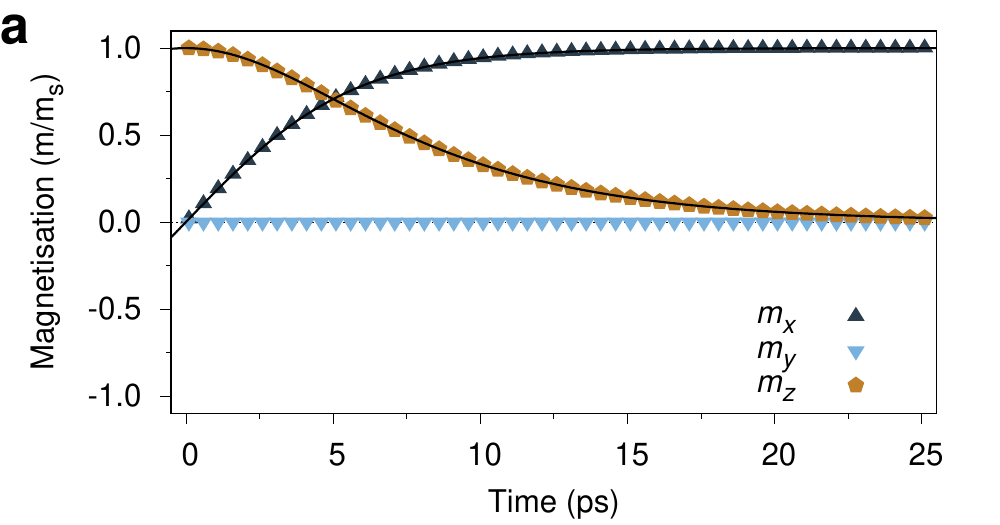}
    \includegraphics[width=\columnwidth]{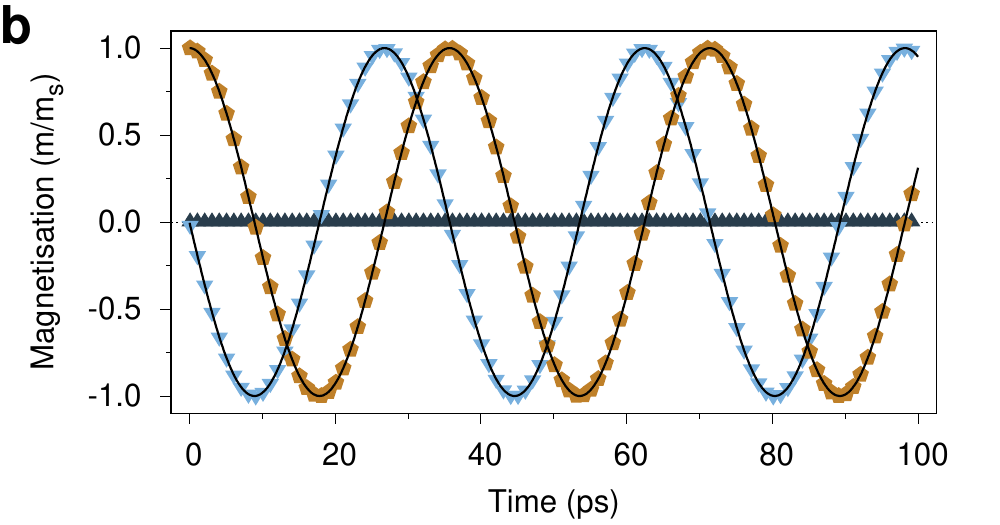}
    \includegraphics[width=\columnwidth]{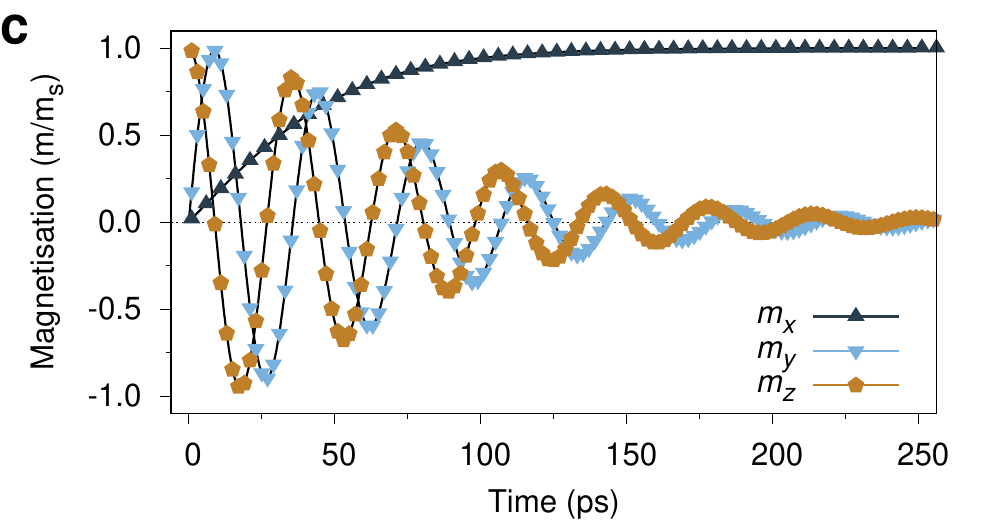}
    \caption{(a) Simulated time-dependent magnetisation for a single magnetic moment under the action of a pure relaxation torque $B_{\mathrm{RT}}^{\mathrm{STT}} = 1.0$ T for initial conditions of $\mathbf{m} = \hat{z}$ and polarization vector $\mathbf{p} = \hat{x}$ (points). Lines show the analytical solution for the time-dependent dynamics. The dynamics show direct relaxation with a characteristic timescale $\tau = 1 / \gamma_e B_{\mathrm{RT}}^{\mathrm{STT}}$. (b) Simulated time-dependent magnetisation for a single magnetic moment under the action of a pure precession torque $B_{\mathrm{PT}}^{\mathrm{STT}} = 1.0$ T showing a continuous precessional motion around the polarisation vector $\mathbf{p}$. (c) Combined dynamics with $B_{\mathrm{RT}}^{\mathrm{STT}} = 0.1$ T and $B_{\mathrm{PT}}^{\mathrm{STT}} = -1.0$ T \new{and zero external applied field} showing standard LLG-like dynamics \new{identical} to a magnetic moment in a constant applied field of $\smBapp = 1.0$ T and $\alpha_{\mathrm{G}} = 0.1$. \new{In general when the relaxation and precessional terms are included with no other terms, LLG-like dynamics are exactly recovered for $\alpha_{\mathrm{G}} = B_{\mathrm{RT}}^{\mathrm{STT}}/B_{\mathrm{PT}}^{\mathrm{STT}}$.}
}
    \label{fig:stt-rp}
\end{figure}

We first consider some simple numerical tests of our method with a single spin model, representative of a single domain magnetic nanodot with magnetization unit vector $\mathbf{m}$. The spin is integrated using the Landau-Lifshitz-Gilbert equation (Eq.~\ref{eq:LLG_STT_solved}) with a Heun numerical scheme\cite{EllisLTP2015} implemented in the \vampire software package\cite{vampireURL,vampire}. The effective field contributions are given by the precession and relaxation torques in Eq.~\ref{eq:STT}. While the physical origins of spin-orbit and spin-transfer torques are different, from a numerical perspective they have the same form and so for simplicity we will only consider a spin transfer torque, but the following section applies equally well to a spin-orbit torque with the same symmetry.

In Fig.~\ref{fig:stt-rp}(a) we plot the dynamics of the magnetization initially along the direction $\mathbf{m} = \hat{z}$ and a polarization vector along $\mathbf{p} = \hat{x}$ under the action of a pure relaxational torque $B_{\mathrm{RT}}^{\mathrm{STT}} = 1.0$ T. In this case we generate the maximum torque (90 degrees between the magnetization and polarization) and expect a pure relaxation motion of the magnetization towards the polarization. In the pure relaxational case this motion is independent of the Gilbert damping $\alpha_G$ since the motion is a pure rotation of the magnetization around the $y$-axis, having explicitly removed the precessional components of the motion. In a similar manner to Hannay~\cite{Hannay,vampire}, the time-dependent magnetisation follows an analytical expression of the form 
\begin{eqnarray}
    m_x(t) &=& \tanh\left(\gamma_e B_{\mathrm{RT}}^{\mathrm{STT}} t\right) \\
    m_z(t) &=& 1 / \cosh\left(\gamma_e B_{\mathrm{RT}}^{\mathrm{STT}}t\right)
\end{eqnarray}
where $t$ is the time. Note here that the analytical solutions are technically approximate since we only apply a first-order correction to the damping component in the Landau-Lifshitz-Gilbert equation when incorporating the spin-transfer torque. However, agreement between the numerical simulation and the analytical form in Fig.~\ref{fig:stt-rp} is excellent. Expressing the relaxation time of the $x$-component of the magnetization in terms of $m_x(t) = \tanh(t / \tau)$, where $\tau = 1 / \gamma_e B_{\mathrm{RT}}^{\mathrm{STT}}$ gives a characteristic relaxation time of $\tau \sim 5.6$ ps and complete relaxation after approximately $3\tau$. Here the intrinsic dynamics of the relaxation is relatively fast, partially due to the large spin-torque field, but also due to the absence of magnetic anisotropy and the fact that the Gilbert damping plays no role in the relaxation dynamics. 

In contrast the dynamics of a pure precessional torque $\Bstp =1.0$ T for the same initial conditions is shown in Fig.~\ref{fig:stt-rp}(b), showing a steady state precession of the magnetization around the polarization direction with period $\tau_\mathrm{p} = \pi \gamma_e \Bstp$. This is fitted to the oscillatory components of the analytical solution of the LLG equation~\cite{Hannay,vampire} given by
\begin{eqnarray}
    m_y(t) &=& -\sin\left(\gamma_e B_{\mathrm{RT}}^{\mathrm{STT}}t\right)\\
    m_z(t) &=& \cos\left(\gamma_e B_{\mathrm{RT}}^{\mathrm{STT}}t\right)
\end{eqnarray}
and is similarly independent of the Gilbert damping, depending only on the magnitude of the spin transfer precession torque $\Bstp$. We note that the sign of the spin-transfer precession torque is important in terms of the sense of rotation of the magnetisation, with negative values being possible, counteracting the usual precessional motion of the intrinsic dynamics~\cite{AbertPRA2017}. The key principle when applying spin torques is reciprocity, i.e. the dynamics of the magnetization in fact represent the dynamics of the incoming itinerant electrons which can be complicated depending on the current density, materials and device geometry. The convention here is that positive values of the precession torque lead to conventional precession of the magnetization, while negative values compete with the intrinsic dynamics. Combining both precession and relaxation torques leads to a standard precession and relaxation of the magnetization towards the incoming polarization direction, as shown in Fig.~\ref{fig:stt-rp}(c) for $\Bstp = -1.0$ T and $\Bstr = 0.1$ T. Note the unusual values here of the precession term being much larger than the relaxation term: conventionally the opposite is true which means that the intrinsic dynamics of spin-induces torques is close to a direct rotation of the magnetization but, depending on symmetry, counteracted by the intrinsic dynamics of the magnetization. Here the Gilbert damping plays a critical role, determining the strength of the counteracting torque from the internal energy contributions of the system. It is universally true that the lower the Gilbert damping the more effective spin-torques are at manipulating the magnetization, with larger Gilbert damping generating larger opposing torques. It is this component that clearly distinguishes the action of spin-torques (which are independent of the Gilbert damping) compared to the intrinsic magnetization dynamics. The intrinsic effects of spin torque are now clear and so in the following sections we consider a range of different problems and device geometries where spin-transfer and spin-orbit torques are applied. 

\subsection{Spin transfer torque switching of a magnetic tunnel junction}
\begin{figure}
    \centering
    \includegraphics[width=\columnwidth]{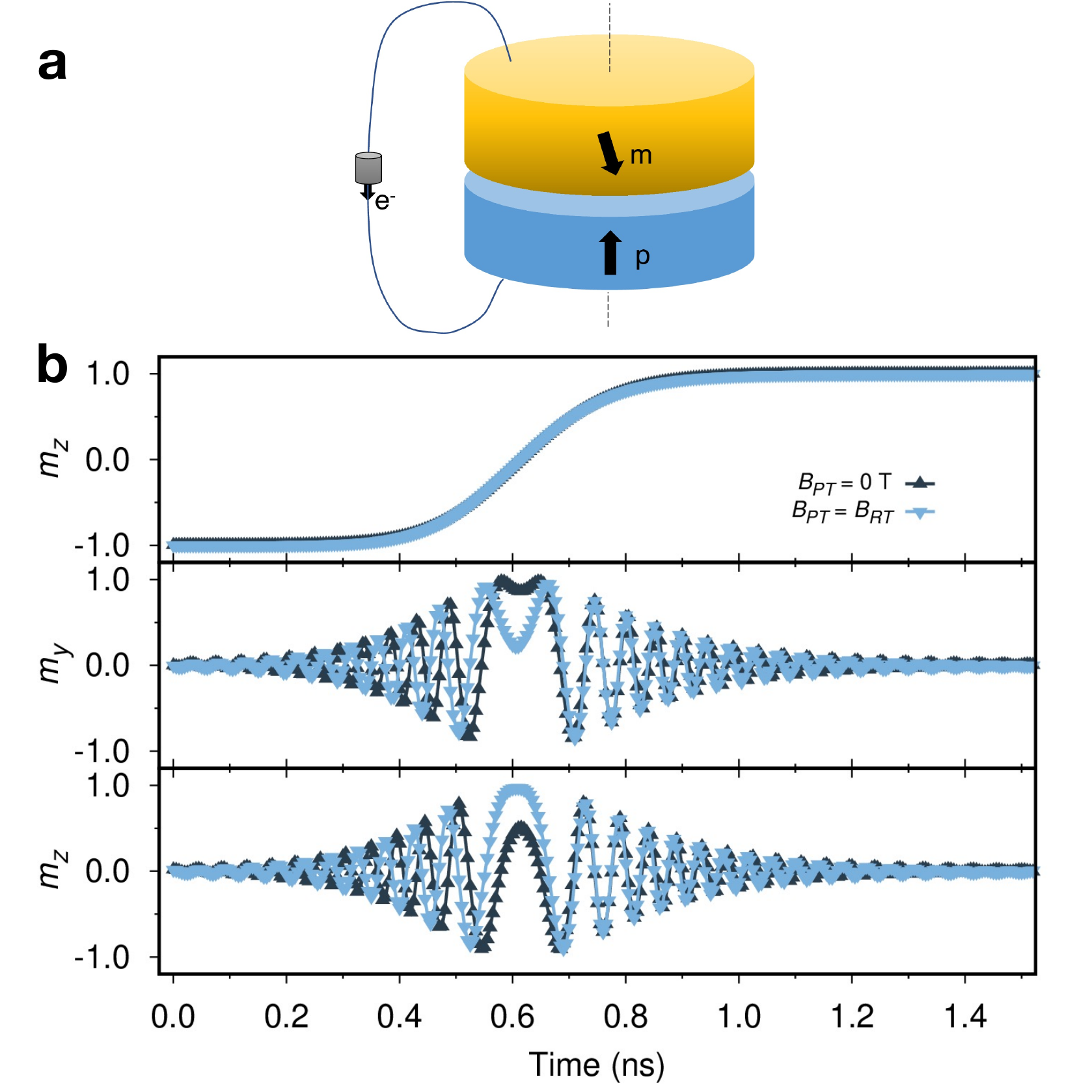}
    \caption{a) Sketch of the simulated MTJ; b) magnetisation dynamics for $B_{\mathrm{RT}}^{\mathrm{STT}}=0.035$ T, $B_{\mathrm{RT}}^{\mathrm{STT}}=0.0$ T (dark blue upward triangles and lines)  and $B_{\mathrm{RT}}^{\mathrm{STT}}=B_{\mathrm{PT}}^{\mathrm{STT}} = 0.035$ T (light blue downwards triangles and lines).}
    \label{fig:STTMRAM_dynamics}
\end{figure}

The classic magnetic tunnel junction (MTJ) was one of the first practical uses for spin transfer torques~\cite{Ikeda2010a} as magnetic random access memory (MRAM). Here we consider a 20 nm cylindrical magnetic tunnel junction in a trilayer structure consisting of two ferromagnets (CoFeB) separated by a thin non-magnetic insulator (MgO). One ferromagnet, the reference layer (RL) has the moment fixed and serves as polarizer for the injected current density along $\bp=\hat{z}$. The other ferromagnet is the free layer (FL) with the magnetisation initially aligned along $-\hat{z}$. To ensure an initial torque can act on the system, the FL magnetisation is canted $1\degree$ from the perpendicular direction. The simulated system is sketched in Fig.~\ref{fig:STTMRAM_dynamics}(a). Given the small dimensions of the system we expect coherent behaviour~\cite{MeoPRB2021,Meo2022} and therefore we can model the system as a single macrospin, representing the FL of thickness 1.3 nm. We model the CoFeB/MgO MTJ \cite{Ikeda2010a} as characterised by $M_s\sim1.3$ T~\cite{SatoPRB2018}, an interfacial anisotropy energy of 1.3 mJ/m$^{2}$, spin-torque efficiency $\eta\sim0.6$ and spin-torque asymmetry $\lambda=\eta^2$. Spin-transfer torque fields of $B_{\mathrm{RT}}^{\mathrm{STT}} = 0.035$ T, $B_{\mathrm{PT}}^{\mathrm{STT}} = 0 $ T corresponding to $j_e=5\times10^{11}$ A/m$^2$ are applied to the FL leading to the characteristic dynamics shown in Fig.~\ref{fig:STTMRAM_dynamics}(b), given by the black lines. 
This result is consistent with previous studies both experimental and theoretical studies, showing typical nanosecond switching timescales. Here our formalism expressing torques as magnetic fields $B_{\mathrm{RT}}^{\mathrm{STT}}$ does not affect the relaxation of the magnetisation nor the the $m_z=M_z/M_s$ component. We verify this by applying a precessional torque $B_{\mathrm{PT}}^{\mathrm{STT}} = 0.035 $ T in addition to the relaxation torque, as shown by the light blue lines in Fig.~\ref{fig:STTMRAM_dynamics}(b). Comparing the two systems, we can see that the different strength of the precessional torque $B_{\mathrm{PT}}^{\mathrm{STT}}$ only affects the precessional dynamics of the magnetisation, whilst $m_z(t)$ is essentially unchanged. By clear separation of relaxational and precessional components of the torque as effective magnetic fields we can more clearly disentangle the two different effects, providing useful insight when interpreting experimental results or performing numerical simulations of switching dynamics.

\begin{figure*}[!tb]
    \centering
    \includegraphics[width=\textwidth]{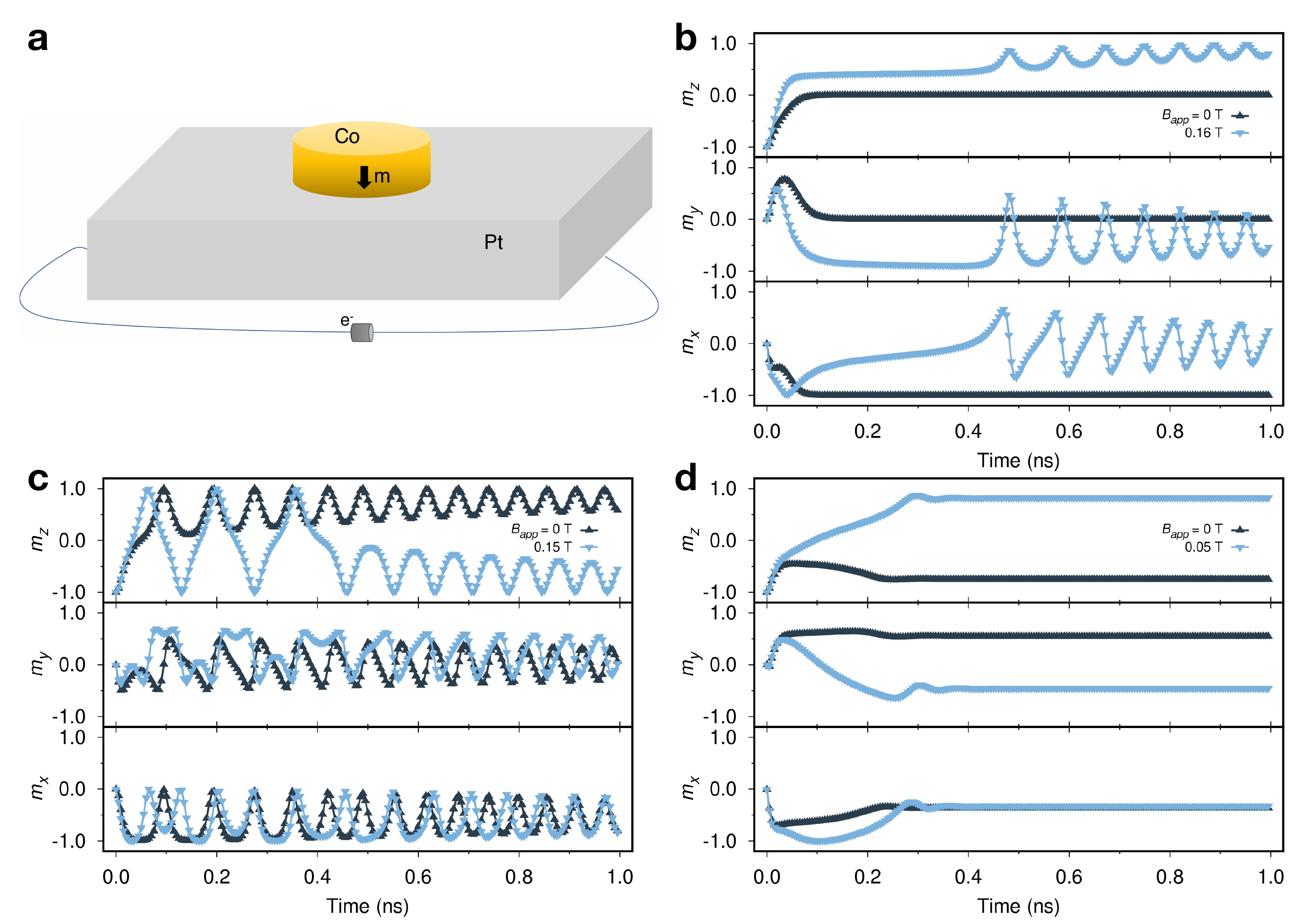}
    \caption{a) Sketch of the simulated Co/Pt system with the Co layer (yellow disk) on top of the heavy metal layer (grey slab) in which the electrical current is injected and induces SHE; b) dynamics of the three reduced components of the magnetisation for $B_{\mathrm{RT}}^{\mathrm{SOT}}=0.4$ T, $B_{\mathrm{RT}}^{\mathrm{SOT}}=0.0$ T, c) $B_{\mathrm{RT}}^{\mathrm{SOT}}=0.0$ T, $B_{\mathrm{RT}}^{\mathrm{SOT}}=0.425$ T and d) $B_{\mathrm{RT}}^{\mathrm{SOT}}=0.4$ T, $B_{\mathrm{RT}}^{\mathrm{SOT}}=0.2$ T. Dark blue upward triangles and lines represent $B_\mathrm{app}=0$ and light blue downwards triangles and lines non-zero $B_\mathrm{app}$.}
    \label{fig:SOT_dynamics}
\end{figure*}

\subsection{Spin-orbit torque switching of a magnetic nanodot}
We now focus on spin-orbit torque (SOT) driven switching of a magnetic nanodot, modelled as a bilayer system Co/Pt, shown schematically in Fig.~\ref{fig:SOT_dynamics}(a). Here we assume a 0.5 nm thick Co cylinder with uniaxial anisotropy lying on top of a Pt contact that is sufficiently thick to assume the spin-orbit torque arises primarily from the spin-Hall effect with a spin-Hall angle of $\thetaSH=0.2$ for Pt~\cite{Zhang2015b}. We inject the current in the Pt along $\hat{y}$ direction, generating a spin polarisation $\sigma$ along the $-x$-direction for 1 ns. Simultaneously we apply a constant magnetic field parallel to the current direction ($\smBapp = B_\mathrm{app}\hat{y}$). 
Initially we consider separate relaxational and precessional torques $B_{\mathrm{RT}}^{\mathrm{SOT}}$ and $B_{\mathrm{RT}}^{\mathrm{SOT}}$ to individually determine the effect of each type of torque. 
Fig.~\ref{fig:SOT_dynamics}(b) shows the dynamics of the Co magnetisation for $B_{\mathrm{RT}}^{\mathrm{SOT}}=0.4$ T and $B_{\mathrm{PT}}^{\mathrm{SOT}}=0$ T, corresponding to $j_e=5\times 10^{12}$ A/m$^2$. The figure compares results obtained for zero field $B_\mathrm{app}=0$ T (black lines) with an applied field of $B_\mathrm{a}=0.16$ T (light blue lines). In this system geometry an external magnetic field is necessary to achieve deterministic switching, since if no \smBapp is applied the magnetisation lies in-plane. In addition, in the absence of an external field, the time evolution of the magnetisation components shows no precessional dynamics, as expected for a pure relaxational torque. By applying an external field, $B_\mathrm{a}=0.16$ T, the switching becomes deterministic. However, given the field-like nature of the torque induced by the applied field, it adds a natural precessional component to the magnetisation dynamics.

In Fig.~\ref{fig:SOT_dynamics}(c) we plot the magnetisation dynamics for a pure precessional torque $B_{\mathrm{PT}}^{\mathrm{SOT}}=0.425$ T and no relaxational torque term. Analogous to the previous case, we compare results for zero field $B_\mathrm{app}=0$ T (black lines) and with an applied field $\smBapp = 0.15$ T (light blue lines). 
The magnetisation shows an initial in-plane re-orientation in less than 100 ps, followed by a precessional motion mainly in the $xz$-plane. 
It is worth noting that $B_{\mathrm{PT}}^{\mathrm{SOT}} = 0.425$ T is the minimum SOT strength that results in the magnetisation exceeding $z=0$. For weaker currents the magnetisation returns to its initial configuration following a similar oscillatory behaviour.
Adding an in-plane applied magnetic field \smBapp affects the precession such that, depending on the strength of the field,  the magnetisation can end in either $+z$ or $-z$ state.
Finally, we include both relaxational and precessional spin-orbit torque terms, setting $B_{\mathrm{PT}}^{\mathrm{SOT}}=1/2 B_{\mathrm{RT}}^{\mathrm{SOT}}$ with $B_{\mathrm{RT}}^{\mathrm{SOT}}=0.4$ T, shown in Fig.~\ref{fig:SOT_dynamics}(d).
For $B_\mathrm{app}=0$ T the magnetisation cannot be reversed and it relaxes in a state close to the initial configuration. We do not observe precessional dynamics and we can conclude that it must oppose the relaxational component of the spin-orbit torque by comparing panels (b) and (d). In the case of both spin-orbit torque components the application of a weak magnetic field of 0.05 T yields deterministic switching of the Co magnetisation. As for $B_\mathrm{app}=0$ T, and differently from the case of pure relaxational torque where a large $B_\mathrm{app}$ was necessary, the magnetisation evolution does not exhibit significant precession.

\subsection{Spin-orbit and spin-transfer torque magnetic random access memory}
\begin{figure}[!tb]
    \centering
    \includegraphics[width=\columnwidth]{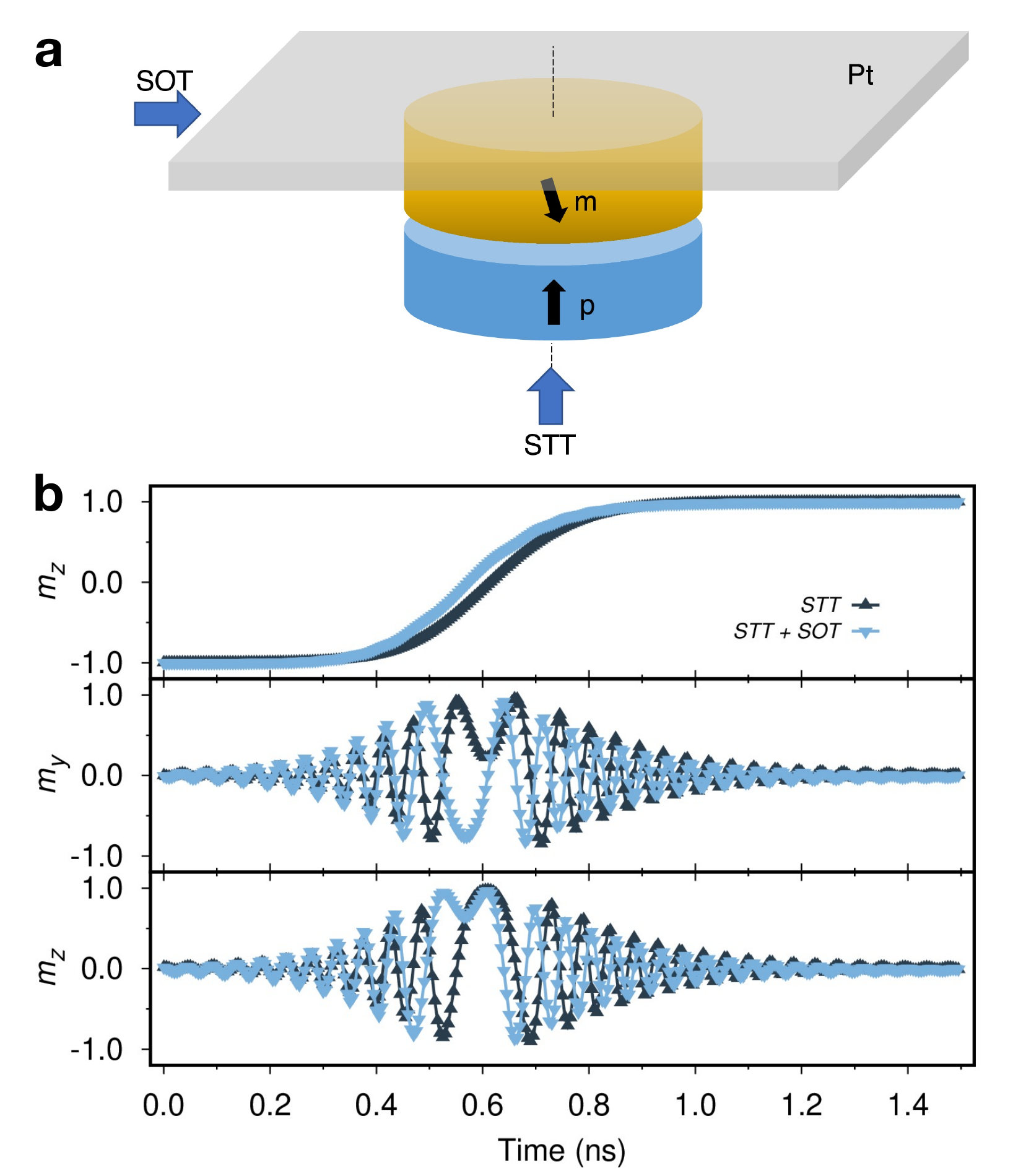}
    \caption{a) Sketch of the simulated MTJ for simulation of combined STT and SOT dynamics, with STT and SOT blue arrows indicating the injection of current for STT (perpendicular to the stack) and SOT (in-plane through the heavy metal) respectively. The free layer is the yellow cylinder, in blue the reference layer and in grey the heavy metal layer; b) dynamics of the three components of the reduced magnetisation comparing the case of pure STT-driven dynamics (dark blue upwards triangles and lines) and combined STT+SOT-dynamics (light blue downwards triangles). 
    $B_{\mathrm{RT}}^{\mathrm{STT}}=B_{\mathrm{PT}}^{\mathrm{STT}} = 0.035$ T and $B_{\mathrm{RT}}^{\mathrm{SOT}}=B_{\mathrm{PT}}^{\mathrm{SOT}} = 0.005$ T when SOT is on.}
    \label{fig:STT_SOT_MRAM_dynamics}
\end{figure}
Recently it has been proposed to combine spin-transfer torque and spin-orbit torque in a single device in order to complement the weaknesses of both \cite{VandenBrink2014,Wang2015,Wang2019,Manchon2019b,Pathak2020,Zhang2021,Meo2022}. 
We consider the same 20 nm cylindrical CoFeB/MgO MTJ previously used to investigate the STT-induced dynamics, where the reference layer is magnetized along $\bp=\hat{z}$, the FL has the magnetisation initially aligned along $-\hat{z}$ and the current density used to generate a spin-transfer torque is injected perpendicular to the MTJ stack.
Differently from the previous works~\cite{Pathak2020,Meo2022}, here we apply the STT and SOT current pulses for the same time given the simplicity of our model and the short time scale considered.
We place a Pt contact on top of the FL to generate spin current via SHE and induce switching in the FL via SOT; this is obtained by injecting a current density along $\hat{y}$--direction. The resultant spin polarisation $\sigma$ is directed  along the $-x$-direction, as in the previously discussed case of pure SOT dynamics. A sketch of the system is presented in Figure~\ref{fig:STT_SOT_MRAM_dynamics}(a).
In order to verify the effect of a combined application of STT and SOT, we will compare the results obtained in this case with those for pure STT-dynamics. For this reason we set the values of the STT fields $B_{\mathrm{RT}}^{\mathrm{STT}} = B_{\mathrm{PT}}^{\mathrm{STT}} = 0.035$ T as done in the pure STT case. 
% In order to verify the effect of a combined application of STT and SOT, we set  $B_{\mathrm{RT}}^{\mathrm{STT}} = B_{\mathrm{PT}}^{\mathrm{STT}} = 0.035$ T as done previously.
% We inject a weak in-plane current through the Pt contact such to generate SOT field of magnitude $B_{\mathrm{RT}}^{\mathrm{SOT}} = B_{\mathrm{PT}}^{\mathrm{SOT}} = 0.005$ T. 
For SOT we assume a weak in-plane current density through the heavy metal that gives field strength $B_{\mathrm{RT}}^{\mathrm{SOT}} = B_{\mathrm{PT}}^{\mathrm{SOT}} = 0.005$ T. 
The magnetisation dynamics resulting from the simultaneous application of STT and SOT current densities is compared with that of pure STT-induced dynamics in Figure~\ref{fig:STT_SOT_MRAM_dynamics}(b). As we can see from this simple example, the combined application of SOT-and STT-dynamics, even for a weak in-plane current density, results in faster switching. This is in agreement with the results reported in literature and it also offers a further verification of our formalism. 
It is worth underlining that such a hybrid device exhibits promising features and can be exploited either to assist STT, as in the case presented here, or to assist SOT ensuring deterministic switching without an external field. The former regime can be utilised to design devices with fast switching on the order of or below nanoseconds with low power consumption, suitable for applications in smart and portable devices. If large SOT current densities are instead injected, the combined dynamics can yield switching on the order of 10 or 100 ps, however this is achieved at the cost of increasing the power consumption~\cite{Meo2022}.

\subsection{Spin-torque nano-oscillators}
\begin{figure}[!tb]
    \centering
    \includegraphics[width=\columnwidth]{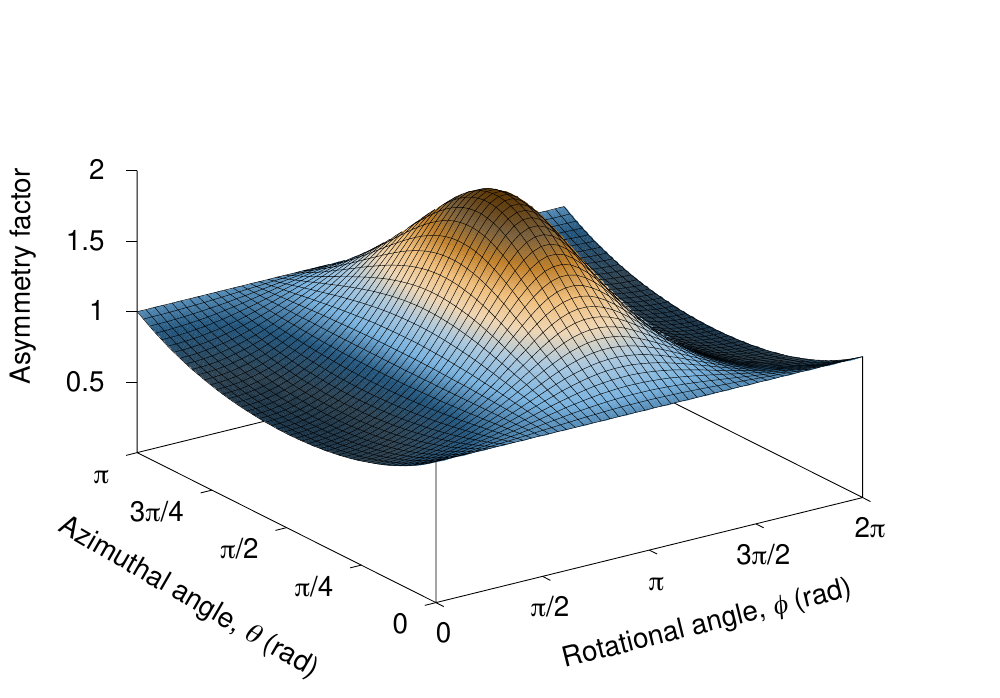}
    \caption{Plot of the spin torque asymmetry $1/(1+\lambda \bM \cdot \bp), \lambda = 0.5$ as a function of the magnetization orientation $\bM$. The asymmetry leads to a rotational variation of the strength of the torque, preventing a natural balancing of torque and a dynamic equilibrium with the magnetization precessing around the $z$-axis.}
    \label{fig:stno-asymmetry}
\end{figure}
Spin-torque nano-oscillators are fascinating devices which can sustain a dynamic precession of the magnetisation under a constant applied voltage. The precession occurs when the spin-transfer relaxation torque exactly compensates the natural magnetic damping of the system, leading to a continuous rotation of the magnetisation. However, the qualitative conditions for oscillation are quite specific and so it is worth exploring these initially. If we consider a magnetic nanodot below the single domain limit at low (zero) temperature then we can model the system as a single macrospin. In the absence of other energy terms, we apply a spin-transfer relaxation torque with an incoming spin polarization along the x-direction, $\mathbf{p} = \hat{x}$, typical for an in-plane polarizer that enables oscillation without the need for an external applied magnetic field. This leads to a direct relaxation of the magnetisation towards the polarisation direction. Adding a perpendicular magnetic anisotropy with easy axis parallel to $\hat{z}$ leads to two possible situations. Above a threshold value of the SORT (spin-orbit-relaxation-torque) the magnetization aligns with the direction of the polarisation $\bp$ in the plane, as the resulting torque from the anisotropy is zero. Below this threshold the magnetization prefers to lie at an angle from the easy $z$-axis in the $y-z$ plane, \textit{perpendicular} to the spin-polarisation direction. This unintuitive behaviour arises due to the balancing of torques. At equilibrium, the torques from all energy contributions are equally balanced, thus the anisotropic torque acts in the direction $\bm \times 2k_u m_z$ while the spin (transfer) torque acts in the direction $\bm \times B_{\mathrm{RT}} \bm \times \bp$, $\bp || x$. As the magnetisation approaches the $x$-direction the spin-torque reduces and so the anisotropic torque remains, since the value of $B_{\mathrm{RT}}$ is too low to align the magnetization with $\bp$ outright. As the magnetisation rotates into the $y$-direction however, the anisotropic and spin torques have opposite signs, and an equilibrium value is found. Thus, for subthreshold values of the spin-torque and the polarization and easy axes are misaligned the magnetization always prefers to align perpendicular to the polarisation direction. An applied magnetic field has a slightly different effect since there is no case where the field and spin-torques are both zero, and so the equilibrium situation is the same as for anisotropy where the magnetization prefers to lie in the $y-z$ plane at some angle (depending on the balance of spin and field induced torques). 

In all of these situations the simple balance of anisotropy and applied magnetic field yields a static equilibrium and so no auto oscillation occurs. The critical parameter is the spin-torque asymmetry, which causes an adjustment of the strength of the spin torque based on the alignment of $\bp \cdot \bm$. Given the functional form in Eq.~\ref{eq:STT_strengthsRT} it is not immediately clear how this arises, and so the angular variation of the spin torque caused by the spin-torque asymmetry $\lambda$ is given in Fig.~\ref{fig:stno-asymmetry}. For the case of a sub-threshold spin-torque (where the magnetization prefers to lie in the $y-z$ plane at some angle less than $\theta = 0^\circ$), the spin-torque asymmetry causes an increase in the spin torque as the magnetization approaches the -$x$-direction ($\phi = \pi)$ which keeps the motion of the magnetization from stopping, leading to a continuous precession of the magnetization. The azimuthal dependence of the spin-torque means that this difference is largest for $\theta = 90^\circ$, causing the magnetization angle $\theta$ to increase and approach the hard axis of the system. However, for the case of zero applied magnetic field the anisotropic torque tends to zero as $\theta = \rightarrow 90^\circ$ and so the precession stops, causing alignment of the magnetization with the polarization direction $\bp$. An applied magnetic field is therefore required to break the symmetry and prevent the total torque approaching zero in the magnetization hard axis. Thus the persistent torque from the applied magnetic field allows for oscillations, while the magnetic easy-axis anisotropy opposes the precession as it provides a torque opposing the spin-torque. An applied magnetic field always breaks the symmetry and so is a general requirement for achieving spin-torque oscillations. This qualitative picture of spin-torque nano-oscillators gives the fundamental ingredients necessary to understand the physical origin of dynamic equilibrium precession in these devices.

\begin{figure}[!tb]
    \centering
    \includegraphics[width=\columnwidth]{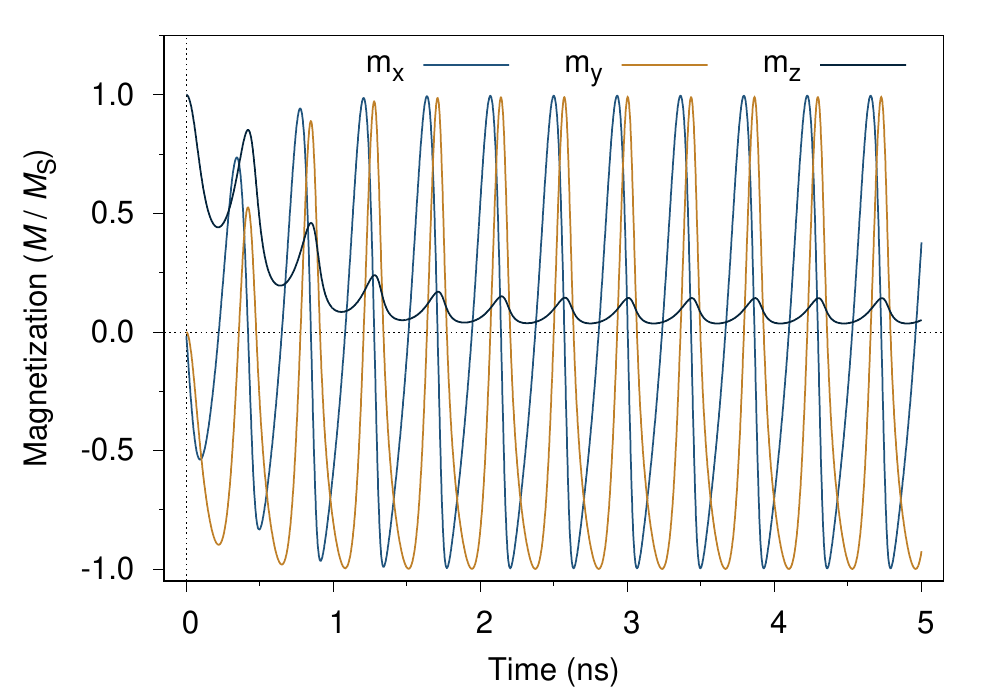}
    \caption{Plot of the dynamic behaviour of a spin -torque nano-oscillator device subjected to a spin-transfer-relaxation-torque of $\Bstr = -0.05$ T, spin-torque asymmetry $\lambda = 0.5$, Gilbert damping $\alpha_{\mathrm{G}} = 0.01$ and external applied magnetic field $B_z = 0.01$ T to break the longitudinal symmetry. The magnetization relaxes from the initial direction $\bM || z$ to a dynamic equilibrium after approximately 2 ns where a continuous oscillation is found.}
    \label{fig:stno}
\end{figure}

In Fig.~\ref{fig:stno} we show the typical dynamics for a nanodot in the single domain approximation subjected to a spin-transfer-relaxation-torque of $\Bstr = -0.05$ T, spin-torque asymmetry $\lambda = 0.5$ and applied magnetic field $B_z = 0.1$ T. The magnetization is initialised along the $+z$-direction. Since the spin-transfer-torque field is less than the applied field strength the equilibrium position is for some intermediate angle of $\theta \neq 0$. Initially the magnetization begins relaxing towards the equilibrium angle, shown by a decrease in the value of $m_z$. However, this decrease is non-monotonic owing to the spin-torque asymmetry, leading to an elliptical precession of the magnetization around the $z$-axis. After two nanoseconds the magnetization has settled into a dynamic equilibrium with a large-angle precession of the magnetization close to the $xy$-plane. However, characteristics of the elliptical precession remain as visible from the non-sinusoidal variation of the $x$ and $y$ components of the magnetization. The magnetization precession is not completely planar but approximately precesses around a virtual field slightly away from the $z$-axis, as seen by the periodic oscillation of the $z$-component of the magnetization. The non-sinusoidal oscillations here are a characteristic feature of the spin-torque asymmetry and large ratio of the spin-torque field and applied field. For smaller values of the spin-torque and large spin-torque asymmetry the precession becomes almost circular.

\subsection{Spin-orbit torque switching of MnPt/Pt bilayers}
\begin{figure}[!tb]
    \centering
    \includegraphics[width=\columnwidth]{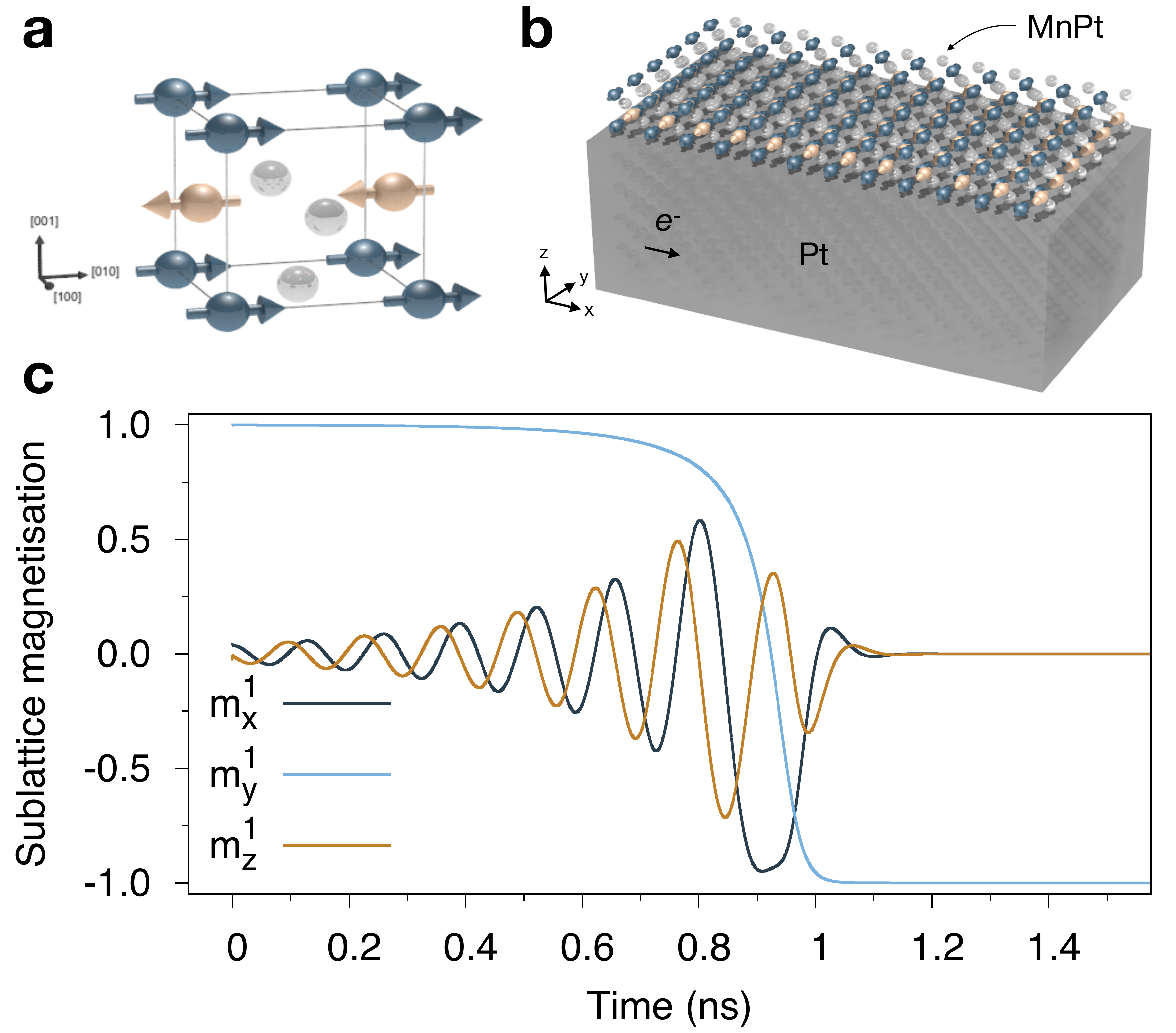}
    \caption{(a) Sketch of the crystal and spin structure of $L1_0$ MnPt, showing a single magnetic sublattice at the MnPt/Pt interface where the spin-orbit torque is largest. (b) Sketch of the simulation setup consisting of a 1nm thick MnPt layer attached to a pure Pt underlayer providing the source for the spin-orbit torque. (c) Time-dependent dynamics of one of the antiferromagnetic sublattices, showing precessional toggle switching with a switching time of around 1 ns.}
    \label{fig:mnptsot}
\end{figure}
Antiferromagnetic spintronics~\cite{MacDonald2011,Gomonay2014,Jungwirth2016,BaltzRevModPhys2018,Jungwirth2018,Jungfleisch2018,Fukami2020} is an emerging field where the sublattice magnetization of antiferromagnets can be directly manipulated by electrical means. This mostly relies on spin-orbit torques although spin-transfer torques are also expected to play a smaller role. At present the theoretical underpinnings and experimental understanding of the dynamics of antiferromagnets are at an early stage, but simulations can assist with understanding the fundamental effects of spin-torque. Unlike ferromagnetic systems, antiferromagnets are ordered at the atomic scale and only weak and coherent excitations can be modelled micromagnetically~\cite{GomonayPRB2010}. We therefore consider an atomistic description of our chosen antiferromagnet, L$1_0$ MnPt, shown schematically in Fig.~\ref{fig:mnptsot}(a). The energetics of the system are described by a spin Hamiltonian~\cite{vampire} of the form 
%----------------------------
\begin{equation}
\mathscr{H} = \sum_{i<j} \smJij \sms_i \cdot \sms_j - \ku \sum_i \smsz^2 
\label{eq:hamiltonian}
\end{equation}
%----------------------------
where $\sms_{i,j}$ are unit vectors describing the directions of local spins $i$ and neighbouring spins $j$, $\smJij$ is the exchange interaction limited to nearest and next-nearest neighbours~\cite{JenkinsPRM2021}, and $\ku = 1.63\times 10^{-24}$ J/atom. The system is evolved using the atomistic Landau-Lifshitz-Gilbert equation~\cite{EllisLTP2015} at $T = 0$ K. The simulated system consists of a thick layer of Pt capped with a 1 nm thick layer of MnPt oriented so that the [010] axis of the crystal lies along the $\hat{y}$ direction. This particular orientation means that the easy axis for the MnPt is parallel to the $\hat{y}$ axis and perpendicular to the current direction along $\hat{x}$, shown schematically in Fig.~\ref{fig:mnptsot}(b). Application of an electrical current along $-\hat{x}$ (with electrons flowing in the direction $+\hat{x}$) gives rise to a spin polarisation along the $-\hat{y}$ direction. The simulated dynamics of one of the two magnetic sublattices is shown in Fig.~\ref{fig:mnptsot}(c) for a spin-orbit relaxation torque field of $\Bsor = 0.1$ T applied to the interfacial Mn layer. Here a small initial angle of $\theta_y \sim 1^\circ$ is given to the antiferromagnet to provide a small initial torque to enable switching as for the ferromagnetic case, otherwise the torque is exactly zero. The system exhibits a rotation of the sublattice magnetization from the $+\hat{y}$ to $-\hat{y}$ direction, while the other sublattice has a corresponding motion to the $+\hat{y}$ direction (not shown). Here the dynamics are somewhat unremarkable, but this is due to the careful consideration of crystal, electronic and magnetic symmetries to yield an example where toggle switching can be achieved. Other more complicated antiferromagnets such as IrMn~\cite{JenkinsPRM2021} and Mn$_2$Au may have significantly different dynamic properties where simulations may play an important role in understanding the switching dynamics.

\section{Conclusions}
In conclusion we have derived a simplified form for the spin-transfer and spin-orbit torques within the explicit LLG equation, where the spin torque is described as an effective magnetic field. This simplifies the numerical implementation with respect to the Landau-Lifshitz equation and removes the mixture of the torque terms that arises from the expansion of the $\alpha_{\mathrm{G}} d\bM/dt$ term. We also propose a nomenclature for the spin torque fields components that relies on the physical origin of the torque (spin-transfer, spin-orbit) as well as the action of the torque on the magnetisation (precession, relaxation). The aim is to provide a more intuitive and clear understanding of the spin-torque processes, and also to enable a simpler interpretation of results in terms of this micromagnetic-like formalism. We have performed numerical tests to validate the approach and to show applications of the proposed formalism. 

\section*{Author contributions}
AM derived the effective field representation of spin torques. AM, CEC and RFLE performed the simulations and analysed the data. RFLE, AW and SJ conceived of and designed the study. AM and RE drafted the paper with contributions from all authors.

%\section{acknowledgments}

%\clearpage
%\bibliography{/Users/rfle500/Documents/Work/Papers/Bibliography/library}
%\bibliography{library,local}
\bibliography{library}
%---------------------------------------------------------------------------%

%\begin{thebibliography}{11}%
%\end{thebibliography}%

\end{document}